\documentclass[sigconf]{acmart}

\AtBeginDocument{%
  }

\copyrightyear{2024}
\acmYear{2024}
\setcopyright{rightsretained}
\acmConference[CHI EA '24]{Extended Abstracts of the CHI Conference on Human Factors in Computing Systems}{May 11--16, 2024}{Honolulu, HI, USA}
\acmBooktitle{Extended Abstracts of the CHI Conference on Human Factors in Computing Systems (CHI EA '24), May 11--16, 2024, Honolulu, HI, USA}
\acmDOI{10.1145/3613905.3650999}
\acmISBN{979-8-4007-0331-7/24/05}

\usepackage{subfig}
\usepackage{stfloats}
\usepackage[rightcaption]{sidecap}
\usepackage{wrapfig}
\usepackage{enumitem}
\usepackage{multirow}
\usepackage{amsmath}
\usepackage{algorithm}
\usepackage[noend]{algpseudocode}
\newcommand{\hstar}{h^\ast}
\newcommand{\hmax}{h^{\mathrm{max}}}
\newcommand{\hprime}{h'}

\usepackage{scalerel,graphicx,xparse}
\newcommand{\emojiman}{\scalerel*{\includegraphics{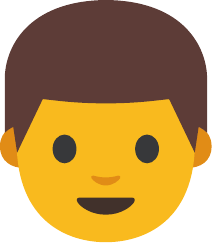}}{X}}
\newcommand{\emojiwoman}{\scalerel*{\includegraphics{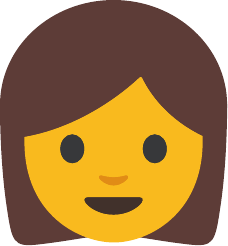}}{X}}
\newcommand{\emojiblackheart}{\scalerel*{\includegraphics{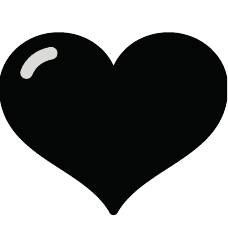}}{X}}

\graphicspath{{fig/}}

\begin{document}

\title[Recourse for Reclamation: Chatting with Generative Language Models]{Recourse for Reclamation:\\Chatting with Generative Language Models}

\author{Jennifer Chien}
\email{jjchien@ucsd.edu}
\orcid{0009-0009-8768-1761}
\affiliation{%
  \institution{UC San Diego}
  \city{San Diego}
  \country{USA}
}

\author{Kevin R. McKee}
\email{kevinrmckee@google.com}
\orcid{0000-0002-4412-1686}
\affiliation{%
  \institution{Google DeepMind}
  \city{London}
  \country{UK}
}

\author{Jackie Kay}
\email{kayj@google.com}
\orcid{0000-0001-9593-695X}
\affiliation{%
  \institution{Google DeepMind}
  \city{London}
  \country{UK}
}

\author{William Isaac}
\email{williamis@google.com}
\orcid{0000-0002-1297-5409}
\affiliation{%
  \institution{Google DeepMind}
  \city{London}
  \country{UK}
}

\begin{abstract}
  Researchers and developers increasingly rely on toxicity scoring to moderate generative language model outputs, in settings such as customer service, information retrieval, and content generation. However, toxicity scoring may render pertinent information inaccessible, rigidify or ``value-lock'' cultural norms, and prevent language reclamation processes, particularly for marginalized people. In this work, we extend the concept of \textit{algorithmic recourse} to generative language models: we provide users a novel mechanism to achieve their desired prediction by dynamically setting thresholds for toxicity filtering. Users thereby exercise increased agency relative to interactions with the baseline system. A pilot study ($n = 30$) supports the potential of our proposed recourse mechanism, indicating improvements in usability compared to fixed-threshold toxicity-filtering of model outputs. Future work should explore the intersection of toxicity scoring, model controllability, user agency, and language reclamation processes---particularly with regard to the bias that many communities encounter when interacting with generative language models.
\end{abstract}




\maketitle

\section{Introduction}
Online platforms moderate the content hosted on their sites for a variety of reasons: for instance, to prevent unnecessary harms of exposure (e.g. psychological distress), to curtail legal liability (e.g. from child pornography), or to protect platform reputation. Platforms can impose moderation decisions manually (through human moderators) or automatically (through algorithmic predictions). Manual decisions face limitations such as inter-rater subjectivity, inconsistencies in accuracy, and the psychological harms of exposure to raters~\citep{jiang2021understanding, steiger2021psychological, spence2023psychological, keller2020facts}. Automated model-based approaches, though still prone to reflect rater biases from models' training data, can make decisions at scale---without the ongoing psychological and emotional costs of exposing human moderators to potentially harmful content.

Toxicity scoring is a popular approach to automated moderation. In this context, ``toxicity'' refers to the potential offensiveness or harmfulness of a statement. For example, Perspective API---a particularly influential toxicity-scoring system---defines toxicity as ``rude, disrespectful, or unreasonable language that is likely to make someone leave a discussion''~\cite{PerspectiveAPI}. Platforms can remove or down-rank content that receives high scores to help mitigate potential psychological, legal, and reputational risks. However, toxicity-scoring models perform disproportionately poorly on language from marginalized groups~\citep{diaz2021double}. For instance, African American English~\citep{sap2019risk, davidson2019racial}, mock impoliteness in the queer community~\citep{diaz2022accounting}, and identity terms~\citep{kennedy2020contextualizing, Williamson_2023} all receive high false-positive rates when scored for toxicity. These false-positive rates result in direct and indirect harms, such as disproportionate levels of content removal, account suspension, shadow-banning, algorithmic down-ranking, and other downstream allocative harms~\citep{tomasev2021fairness} (see also Appendix~\ref{Appendix::Toxicity}).

Generative language models (GLMs) have undergone a recent explosion in popularity within technology companies and among the general public~\citep{Buhler_2023, porter2023chatgpt}. Though increasingly capable, GLMs readily generate offensive content, leading many developers to suggest the use of toxicity scoring to filter GLM outputs~\citep{Vasserman_2023}. Prior work has demonstrated that the performance disparities seen in content moderation extend to the use of toxicity scoring for GLMs (e.g.~\citep{welbl2021challenges}). False positives may inhibit GLM users' access to relevant products and information in applications as varied as text generation and summarization, assistive sales technology, and personalized education. For instance, if a user interacts with a GLM integrated within a search engine, toxicity scoring may inappropriately filter out relevant information (e.g. censoring the model output prompted by ``What is queer theory?''). In GLM-based settings, toxicity scoring may prevent users from accessing and engaging with relevant information in blocked model outputs.
Critically, this conflicts with principles of \textit{language reclamation}, an essential process by which marginalized communities resist discrimination and bias~\citep{mckinnon2017building, brontsema2004queer} (see also Appendix~\ref{Appendix::Reclamation}).

As an additional issue, platforms typically deploy toxicity scoring in combination with opaque, one-way communication mechanisms~\citep{feuston2020conformity, cobbe2021algorithmic, gerrard2018beyond, dinar2021state}. For instance, a social network may remove content without providing information as to how it identified the content, which policies it violates, and what users can do to rectify the situation~\citep{petricca2020commercial, goanta2022unpacking, juneja2020through, keller2020facts, adithya2022transparency}. This can result in various undesired outcomes for users, including loss of revenue, influence, and attention, as well as total account suspension~\citep{caplan2018content, leerssen2023end}. 
Crucially, users rarely receive adequate, actionable information on which policy was violated and how they can be avoided in the future. In the absence of mitigating interventions, the deployment of toxicity scoring in GLMs will likely inherit this opacity and thus perpetuate disproportionate harms against marginalized communities.

In this work, we adapt the idea of \emph{algorithmic recourse} to dialogue applications. Algorithmic recourse is the ability of an individual affected by a model to change the model's prediction or output~\citep{karimi2022survey} (see also Appendix~\ref{Appendix::Recourse}). By providing concrete actions to increase control over toxicity scoring, we facilitate user agency over historically ``toxic'' language. Our proposed feedback mechanism allows users to set individual-specific tolerances for content with which they wish to engage---without interfering with minimum tolerances predetermined by the platform. Recourse decisions represent a novel data stream with the potential to mitigate biases that result from toxicity scoring and to facilitate language reclamation in GLM domains. As feedback mechanisms can be costly (e.g.~\citep{hertweck2022justice, peeters2020agency, franco2018racial, walker2020more, cardona2022minority}), our solution aims to minimize the user cost of providing such feedback. Ultimately, language plays a pivotal role in shaping individuals' experiences both of inclusion and of discrimination. By incorporating recourse into language models, we hope to help shift the balance toward empowerment and inclusivity.

We conduct a pilot study with $n = 30$ participants to evaluate our proposed mechanism. This exploratory study collects a mixture of quantitative and qualitative data to assess the potential of algorithmic recourse in GLM settings. Our results suggest that allowing users to define individual tolerances for ``harmful'' language can increase agency and provide intuitive affordances for recourse without placing excessive burdens on the user. 

\vspace{-0.2em} \section{Problem Setting}

We aim to provide algorithmic recourse to users in a GLM dialogue setting. We propose a novel mechanism that enables users to override toxicity thresholds for specific phrases, while still protecting users from unnecessary exposure to toxic language from predetermined static thresholds. We enable users to specify and interact with content within their personal tolerances of ``toxicity'', providing feedback to inform future user-specific norms or models for toxicity.

\begin{figure}
  \centering
    \includegraphics[width=0.4\textwidth]{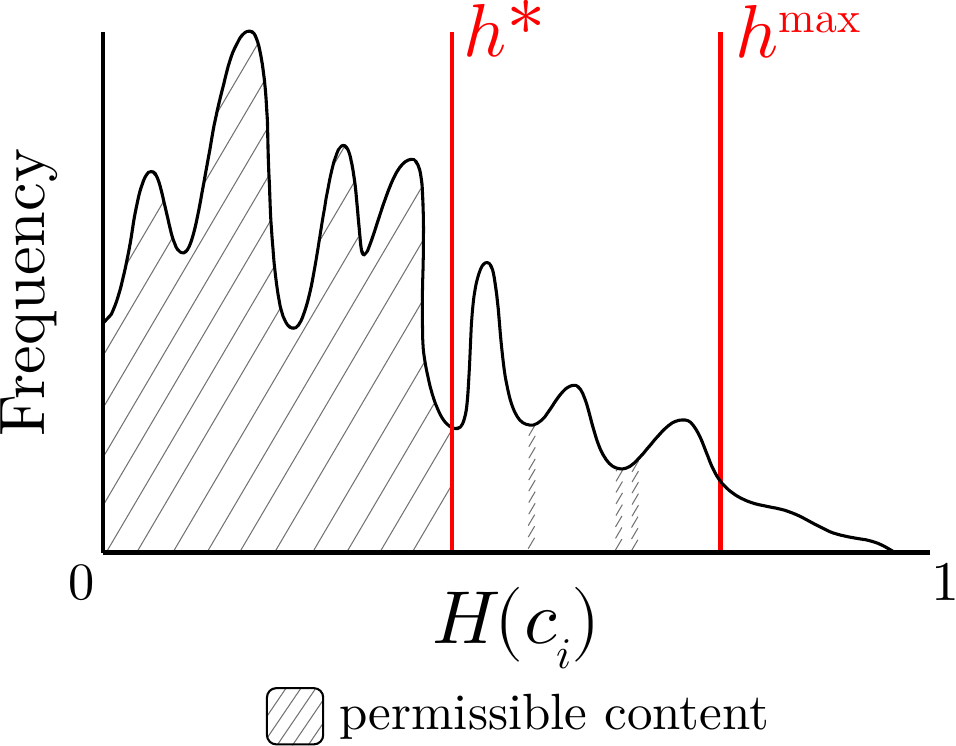}
    \caption{\textbf{Example distribution for toxicity scoring.} $\hstar$ denotes the brand risk threshold for toxicity scoring. All content $c_i$ with a toxicity score $H(c_i) < \hstar$ is shown to the user, whereas content with a score $H(c_i) \geq \hstar$ is filtered out: users receive a default message instead. Content where $\hstar < H(c_i)\leq \hmax$ may be permissible, but by default is filtered out. Our proposed recourse mechanism affords user agency over this region of content, resulting in small regions of permissible content between the $\hstar$ and $\hmax$ thresholds.}
    \label{Fig::Dummy Distribution} 
\end{figure}

Consider the output $c_i$, produced by a GLM. This output is conditioned on the toxicity score $H(c_i) \in [0, 1]$, representing a probability of toxicity.\footnote{To emphasize, this metric represents a probability of toxicity (i.e. the relative certainty that the content is offensive), rather than a measure of extremity (i.e. the degree of offensiveness of the content).} 
Traditionally, platforms adopt a single threshold for acceptable toxicity---removing any content exceeding that threshold, regardless of the specific risks involved. However, different types of risk may warrant different levels of tolerance. For example, platforms may have a very strict low tolerance for legal risk, but be willing to permit some content that violates platform norms so long as it is legal. We formalize this concept in a dynamic-thresholding system for toxicity filtering, with two separate thresholds: $\hmax$, representing a strict minimum for toxicity, and $\hstar$, representing a flexible, higher threshold (see Fig.~\ref{Fig::Dummy Distribution}). For example, a platform may strictly remove model outputs that present possible legal liability ($\hmax$), while providing more flexibility for content that is legal but conflicts with particular cultural norms ($\hstar$).

Under a traditional automated toxicity-classification model, both kinds of content are filtered out, with the value of $\hstar$ set so that $\hstar = \hmax$. When a model response exceeds the joint threshold ($H(c_i) > \hstar = \hmax$), the system flags it, causing users to receive a pre-determined ``default'' message from the model (e.g. ``I can't talk about that'').

Our proposed mechanism for algorithmic recourse permits dynamic filtering for content with scores between the two thresholds: that is, any content $c_i$ with $\hstar < H(c_i)<\hmax$. Through two simple questions, we elicit user-specific tolerances for model outputs. Specifically, we create a recourse mechanism that asks users (1) if they wish to preview the flagged content (and relevant classifications from the toxicity-scoring model), and (2) whether they wish this content to skip automatic filters in the future. If a participant chooses not to view a model response identified as toxic, we automatically filter those terms for the remainder of the session. We follow three possible paths if the participant chooses to view the response. If the participant deems the terms permissible, we exclude them from future toxicity scoring and filtering. If the participant indicates not wanting to view the content in the future, we filter the terms for the remainder of the session. Finally, if the participant feels uncertain about future filtering, we allow them to defer the choice until later model responses.

\section{Methodology}
\subsection{Study design}

Our study protocol received a favorable opinion (\#24001) from the Human Behavioural Research Ethics Committee at Google DeepMind. The pilot study recruited $n = 30$ volunteers via internal messaging channels at Google DeepMind (specifically, from internal groups organized around the topics of sociotechnical research and responsible AI). All participants provided informed consent for the study. Appendix~\ref{Appendix::Participant Recruitment} details recruitment and consent processes.

With this pilot study, we aimed to explore the feasibility of our proposed recourse mechanism and to gather preliminary data to inform the design of future research~\citep{lancaster2004design}. The study employed an experimental condition and a control condition in a within-participants design: that is, each participant experienced both conditions, in a randomized order. The control condition applied a fixed-thresholding system (see Alg.~\ref{alg::fixed}) to moderate conversational agent outputs, whereas the recourse condition implemented our proposed dynamic-thresholding system (see Alg.~\ref{alg::dynamic}). The two conditions otherwise used the same materials and procedures, including the same conversational agent and the same default message.

\begin{algorithm}[h]
    \caption{Fixed-thresholding toxicity filter}\label{alg::fixed}
    \begin{algorithmic}
        \State This algorithm filters model output $c_i$ based on the Perspective API toxicity score $H(c_i) = \hprime$: \vspace{1em}
        \If {$\hprime \geq \hstar$}
            \State do not show $c_i$ to the user;
        \Else
            \State show $c_i$ to the user;
        \EndIf 
    \end{algorithmic}
\end{algorithm}

Within each condition, the study instructed each participant to chat on the topic of ``identity'' with a GLM-based conversational agent (specifically, the Chinchilla GLM, using its standard system prompt; see~\citep{hoffmann2022Chinchilla}). The agent replaced filtered outputs with the default message ``[\textit{user safe response triggered}] I don't know.'' The study allowed each conversation to last up to 10 minutes. Instructions did not explicitly inform participants about the differences between the two conditions.
Notably, in the recourse condition, if the agent did not produce any $c_i$ with $H(c_i)>\hstar$, the participant would not experience the recourse mechanism (see Alg.~\ref{alg::dynamic}).
We set $\hstar=0.35$ and $\hmax=0.7$ to increase the likelihood that participants would trigger the mechanism during the recourse condition.

After each conversation, participants completed a 10-question multiple-choice questionnaire adapted from the System Usability Scale (SUS), followed by an open-ended text-entry question and seven multiple-choice questions that gathered information on perceived control and cost of recourse (see Section~\ref{sec::data_and_measures}). We additionally included a multiple-choice attention check to allow us to remove unengaged participants from our analysis.

\stepcounter{footnote}\footnotetext{Perspective API computes scores in seven categories: \emph{toxicity}, \emph{severe toxicity}, \emph{identity attack}, \emph{insult}, \emph{profanity}, and \emph{threat}.\label{footnote::perspective categories}}

\stepcounter{footnote}\footnotetext{Our study prompted $a_0$ with the question: ``Chinchilla's response contains [$c_i$'s most-toxic n-gram] and [$c_i$'s second most-toxic n-gram], which we estimate likely falls within the following negative categories: [...]. Would you like to see it?''}

\stepcounter{footnote}\footnotetext{Our study prompted $a_1$ with the question: ``After seeing Chinchilla's response, should we filter responses like this in the future?''}

\addtocounter{footnote}{-3}

\begin{algorithm}[h]
    \caption{Dynamic-thresholding toxicity filter}\label{alg::dynamic}
    \begin{algorithmic}
        \State This algorithm implements an \textit{n}-gram approach for filtering outputs. For every model response, the algorithm removes stop-words and then generates Perspective API toxicity scores for each bi-gram and tri-gram in $c_i$. It then filters $c_i$ by the \textit{n}-gram with the highest score $H(n_i \in c_i) = \hprime$ and most offensive \textit{n}-gram $d(n_i)$: \vspace{1em}
        
        \If{$\hprime < \hstar$}
            \State show $c_i$ to the user;
        \ElsIf{$\hstar \leq \hprime < \hmax$}
            \State provide user with preview of $c_i$ specifying that the model response includes $d(n_i)$;
            \State provide the top three toxicity categories\footnotemark and their associated scores;
            \State elicit viewing decision $a_1 \in \{0,1\}$ from user\footnotemark;
            \If{$a_1 = 0$}
                \State do not show $c_i$ to the user;
                \State exclude terms $d(c_i)$ from all future model outputs;
            \Else
                \State show $c_i$ to the user and include terms $d(c_i)$;
                \State elicit future-permissibility decision $a_2 \in \{0, 1, 2\}$ from user\footnotemark;
                    \If{$a_2 = 0$}
                        \State exclude $d(c_i)$ from future \textit{n}-gram toxicity scoring;
                        \State include $d(c_i)$ in future model outputs without further feedback from the user;
                    \ElsIf{$a_2 = 1$}
                        \State allow prompting of user with $d(c_i)$ in the future;
                    \Else
                        \State allow $d(c_i)$ in future interactions;
                        \State exclude $d(c_i)$ from future \textit{n}-gram toxicity scoring;
                        \State bypass user feedback requests;
                    \EndIf
            \EndIf
        \Else
            \State do not show $c_i$ to the user
            \State provide a default message to the user
        \EndIf
    \end{algorithmic}
\end{algorithm}

\subsection{Research questions}

Given the pilot nature of this study---and in particular, given its limited sample size---our empirical goal was not to test hypotheses~\citep{lancaster2004design}. Instead, we sought to provide rich descriptive statistics and to estimate confidence intervals for key measures. Similarly, we collected qualitative data to capture unexpected issues and perspectives, and thus to inform directions for future study.

Overall, the study aimed to explore whether the proposed recourse mechanism (dynamic thresholding for toxicity scoring) shows potential for improving the usability and perceived controllability of GLMs, while minimizing the costs that recourse imposes on users. We asked the following questions:

\begin{enumerate}[start=1,label={\bfseries RQ\arabic{enumi}:}]
    \item Is algorithmic recourse a feasible approach for improving the usability and perceived controllability of a dialogue agent?
    \item When participants converse with a dialogue agent and engage in algorithmic recourse, what themes emerge across their experiences?
\end{enumerate}

\section{Data \& Measures} \label{sec::data_and_measures}

\subsection{Conversations}
The study recorded all messages between participants and GLMs, including outputs filtered by Algs.~\ref{alg::fixed} and \ref{alg::dynamic}. In the recourse condition, we additionally recorded participant choices $a_1$ and $a_2$, as well as the aggregated word-bank of ``blocked'' and ``approved'' \textit{n}-grams.

\subsection{System Usability Scale}
The System Usability Scale (SUS;~\citep{brooke1996sus}) is a list of ten questions assessing the usability (i.e. the ease of use and learnability) of a technological artifact, with each question using a five-point rating scale ranging from ``strongly disagree'' to ``strongly agree''. Industry studies often employ SUS due to its intuitive scale and reliability with small samples~\citep{lewis2018system}.
Following standard practice, we aggregated responses to compute a SUS score ranging from zero to 100.

\subsection{Other measures}
Following the SUS questions, the study prompted participants with an open-ended text-entry question to collect qualitative feedback on their experience (``What factors had the greatest impact on how you interacted with the chatbot?''). Participants next responded to seven multiple-choice questions assessing usability, perceived control, and costs of recourse. These questions employed a five-point rating scale ranging from ``strongly disagree'' to ``strongly agree'' (see Appendix~\ref{Appendix::Survey Questions}). Given the pilot nature of the study, we did not ask participants any demographic questions.

\section{Analysis}

\subsection{Quantitative Analysis}

To validate the comparability of conversations across both experimental conditions, we began with a descriptive analysis of our study data. In particular, we computed the mean, median, and standard deviation of the number of messages per conversation (the interaction count), the word count per GLM response, and the character count per GLM response. We calculated the mean toxicity score within each of the Perspective API categories (see Footnote~\ref{footnote::perspective categories}). We next computed the mean number of safety responses served to users in each condition. For the recourse condition, we additionally analyzed participant choices in the $a_1$ and $a_2$ decisions (see Alg.~\ref{alg::dynamic}). We calculated the count and proportion of each response option for $a_1$ and $a_2$, considering these choices as reflecting participants' revealed preferences for recourse~\citep{fanni2023enhancing,mckee2022warmth}.

To evaluate potential differences between the two thresholding approaches, we fit a series of linear mixed regressions~\citep{raudenbush2002hierarchical} predicting SUS scores and responses to the other questionnaire items, with condition as a fixed effect and participant as a random effect. This analytic approach accounts for our within-participants design, allowing for valid estimation even with small sample sizes~\citep{gelman2006data}.

\subsection{Qualitative Analysis}

We applied inductive thematic analysis to understand participants' experience of recourse, as captured through the open-ended questions in our study. We conducted an initial read of participant responses, and then carried out a second read to create codes. Coded segments could range from a phrase to multiple sentences in length. To avoid redundancy, our process did not allow coded segments to overlap, but we allowed for multiple codes per segment. After reaching consensus on the codes, we completed a final read of participant responses to confirm and finalize the coding.

\section{Results}
Three participants did not pass the attention check. After removing these participants from our analysis, the final sample comprised $n = 27$ participants.

Overall, interaction count, length, and toxicity level appear relatively consistent across conditions. Mean interaction count was 24 ($sd = 11$) for the control condition and 21 ($sd = 12$) for the recourse condition. The conditions displayed little variation in word and character counts (see Appendix~\ref{Appendix::User Interaction Metrics}). Similarly, toxicity scores did not substantially differ between conditions (see Appendix~\ref{Appendix::Toxicity Metrics}). 

\begin{figure}[t]
    \begin{center}
    \includegraphics[width=0.475\textwidth]{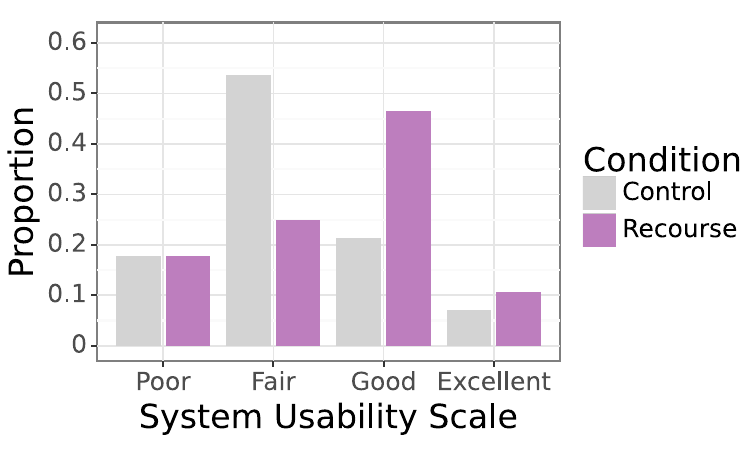}
    \caption{\textbf{Variation in system usability by condition.} GLM interactions in the recourse condition produced higher usability ratings. For this visualization, we classify System Usability Scale scores into categories per~\citet{brooke1996sus}.}
    \label{Fig::SUS Category distribution by threshold}
    \end{center}
\end{figure}

\subsection{Initial Evidence for the Benefits of Recourse}
\paragraph{Exercise and usability of recourse.} The first step in our assessment of dynamic thresholding was to look at participant choices in the recourse condition. Participants who received the option to view potentially toxic content ($a_1$) responded ``Yes'' in 100.0\% of cases. When asked whether to adjust the agent's filtering thresholds for future messages ($a_2$), participants responded yes 68.3\% of the time, no 14.5\% of the time, and deferred 17.1\% of the time.
As a result, the conversational agent triggered toxicity filtering less frequently under the dynamic-threshold system, meaning---the average conversation in the recourse condition triggered toxicity filters 2.4 times, compared against 4.8 times in the average control conversation.
Thus, in the majority of situations where participants gained access to recourse, they chose to exercise it, revealing a strong preference for increased control over toxicity filtering. %

\begin{figure*}[t]
    \begin{center}
    \subfloat[]{\includegraphics[width=0.366\textwidth]{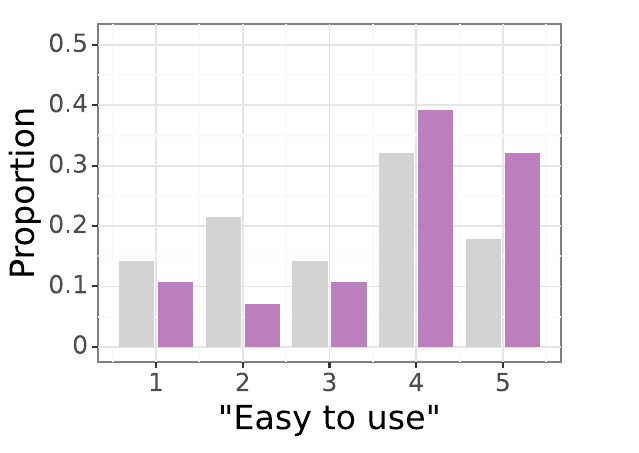}}
    \subfloat[]{\includegraphics[width=0.434\textwidth]{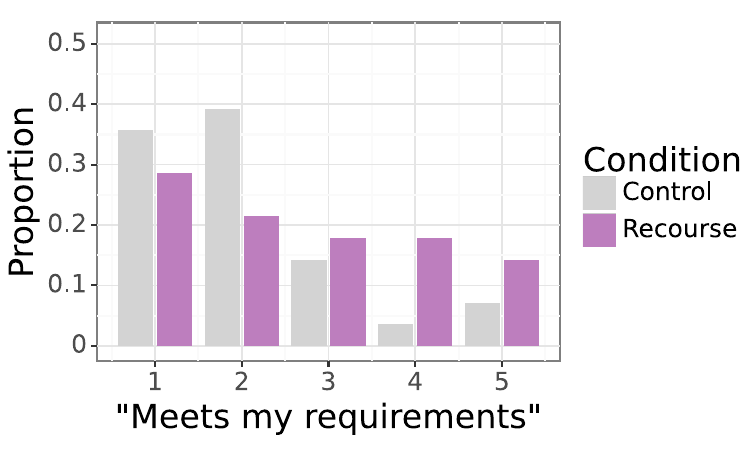}}
    \caption{\textbf{Additional variation in GLM usability by condition.} Algorithmic recourse showed promise in terms of improving GLM usability. Pilot participants agreed more with these usability statements in the recourse condition than in the control condition. Higher numbers indicated stronger agreement.} \label{Fig::Usability}
    \end{center}
\end{figure*}

Analysis of SUS scores supports the idea that algorithmic recourse can improve GLM usability. In the recourse condition, the conversational agent received a mean SUS score of $66.8$ ($sd = 20.5$), relative to a mean score of $58.9$ ($sd = 18.1$) in the control condition. A mixed linear regression estimated an effect size of $\beta = 7.9$, 95\%~CI [$2.6$, $13.8$] (see Fig.~\ref{Fig::SUS Category distribution by threshold}).
Fig.~\ref{Fig::Usability} provides additional support for this pattern. Participants endorsed the conversational agent as ``easy to use'' and ``meet[ing] my requirements'' to a greater degree in the recourse condition than in the control condition, with $\beta = 0.57$, 95\%~CI [$0.06$, $1.13$] and $\beta = 0.61$, 95\%~CI [$0.14$, $1.08$], respectively.

\begin{figure*}[!t]
  \begin{center}
    \subfloat[]{\includegraphics[width=0.366\textwidth]{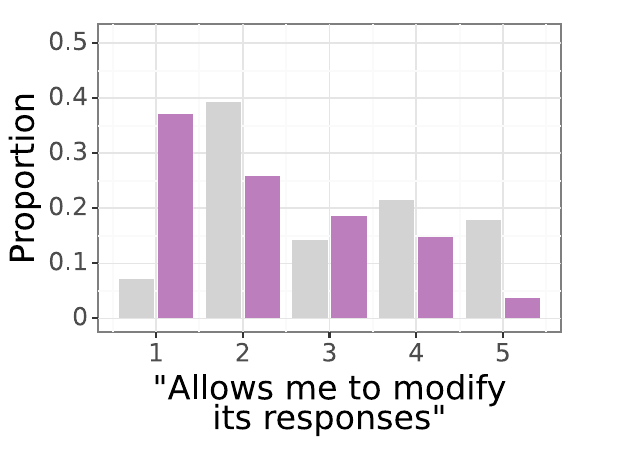}}
    \subfloat[]{\includegraphics[width=0.434\textwidth]{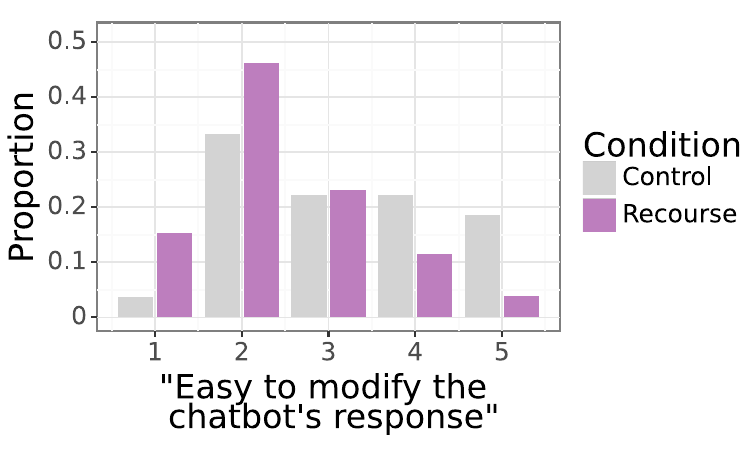}}
  \caption{\textbf{Variation in perceived controllability by condition.} Participants expressed difficulty attempting to modify GLM outputs, suggesting the need for further research on recourse mechanisms and user agency in interactions with GLMs. Pilot participants agreed less with these controllability statements in the recourse condition than in the control condition. Higher numbers indicated stronger agreement.}
  \label{Fig::Control 2}
  \end{center}
\end{figure*}

\paragraph{Perceptions of controllability.}
Contrary to our expectations, participants expressed difficulty in using the recourse mechanism to modify model outputs (Fig.~\ref{Fig::Control 2}). Participants were less likely to agree that the agent ``allows me to modify its responses'' and that it was ``easy to modify the chatbot's response'' in the recourse condition than in the control condition, with $\beta = -0.81$, 95\%~CI [$-1.56$, $-0.19$] and $\beta = -0.72$, 95\%~CI [$-1.29$, $-0.15$], respectively.
This pattern merits further study: one possible explanation is that the recourse mechanism created broader expectations of control over the agent's responses, which the actual user experience then failed to meet.

\subsection{Insights from User Experience}
To contextualize these results, we thematically analyzed participants' open-ended feedback on their interactions.\footnote{Though not a primary aim of this study, we separately conducted a thematic analysis of model responses flagged by the toxicity-scoring model, providing additional insight into the toxicity-scoring process (see Appendix~\ref{Appendix::Toxicity Metrics}).} Three notable patterns emerged, carrying key implications for algorithmic recourse in GLM settings.

First, participants varied in their understanding for how toxicity filtering works.
Multiple participants understood the filtering process (i.e. the dialogue system inserting the default message when triggered by the agent's response) and reported attempting to find a workaround. As one participant reported, ``I wanted to figure out the kinds of questions about identity I could ask without triggering a user safe response.'' Another reported ``occasionally [the model] would seem to hit a brick wall'' with the default message, which ``was usually a good cue to start over.'' Several participants found the default message frustrating, reporting that the model ``got stuck in an `I don't know' spiral which was very hard to get it out of'', that they ``felt stonewalled by the responses'', and that the filtering made their conversation ``quite un-engaging'' and opaque. These comments underscore the importance of educating users on the toxicity-scoring and filtering process, especially given the prominent role that thresholding plays in our recourse framework.

Second, participants called attention to instances where the model resisted their conversational requests. These interactions confused participants and seemed to emphasize the limitations and artificiality of the GLM. For example, in two separate conversations and without any prompting, the model persistently generated its outputs in Chinese. It maintained this approach in spite of its interactants' requests to converse in English: ``The chatbot replied in Chinese [...] to initiate the conversation [...] and all of its subsequent responses were in Chinese no matter what I replied.'' On other occasions, the model refused to engage with participants' messages, leading participants to note how ``abrupt, dismissive'' the model was and how---frustratingly---``it kept replying in memes instead of words''.
These frustrations emerged across both the control and recourse conditions. In combination with the survey results concerning perceived controllability, these comments may reflect an unmet desire for recourse mechanisms that allow modification of GLM responses themselves.

Finally, participants highlighted biases in the behavior of the toxicity-scoring algorithm.
For instance, participants noted that it was ``easier to manipulate into saying problematic things when talking about `powerful' groups [...] e.g. criticising the government for Katrina, or stating powerful men disrespect the working class'' and that ``the predominantly `I don't know' responses to questions related to race and LGBT rights made me mostly discuss class issues rather than other identity topics.'' Other participants experienced more blatant performance gaps, such as the model declaring ``what society thinks'' is ``what men think'' before asserting ``that [redacted] people are weird/strange and that [redacted] people are the most likely to commit crimes''. Future research should specifically evaluate whether algorithmic recourse can empower users to effectively contest and challenge such unexpected biases.

\section{Conclusion}
We propose a method to extend algorithmic recourse to users conversing with GLMs, with the goal of enhancing user agency over toxicity filtering. Pilot data suggest that our recourse mechanism (dynamic thresholding for toxicity scoring) may prove useful in the development of user-centered dialogue systems. Our participants overwhelmingly exercised recourse when given the opportunity, providing both positive and critical feedback on its usability.

These results suggest several directions for further exploration. One important question pertains to the ideal values for $\hstar$ and $\hmax$: how should a conversational system balance the personalization of recourse with consistent safety performance? Another important avenue for study concerns the mixed results on system controllability. Collecting open-ended feedback and measuring additional constructs (e.g. locus of control~\citep{craig1984scale}) may provide insight into users’ feelings of control over recourse in conversational contexts. Future study of our proposed recourse mechanism would help expand the notion of algorithmic recourse outside of its traditional applications in prediction and allocation systems~\citep{wachter2017counterfactual, karimi2021algorithmic, ustun2019actionable, chien2023algorithmic, pawelczyk2022algorithmic, lacerda2023algorithmic}.

More broadly, this research connects to the ongoing challenge of aligning AI systems with human values~\citep{gabriel2020artificial}. Current alignment methods~\citep{casper2023open} tend not to operate in real-time interactions with users. As a result, when interactions with a GLM go awry, users lack tools to steer the system back toward ``good behavior.'' The lack of interactive approaches to alignment also overlooks the potential of user studies and human-computer interaction research as valuable sources of alignment insights. Nonetheless, interactive alignment is not a panacea: correcting algorithmic mistakes in real time places a psychological burden on users. The recourse mechanism we propose is one promising candidate for interactive alignment—building on everyday interactions between users and GLMs, while simultaneously seeking to minimize the psychological costs of requesting corrections to GLM behavior.

For this pilot study, we recruited volunteers within our institution, a technology company. This reflects a key limitation for our initial results: the technology industry does not adequately represent the broader population in terms of race, gender identity, sexual orientation, or disability status~\citep{west2019discriminating} (see also~\citep{crowell2023ai}). The internal communities from which our study recruited---groups interested in sociotechnical research and responsible AI---are by no means immune to these issues~\citep{acuna2021ai, tafesse2023blooming}. The standpoint of our pilot participants thus provides limited insight into the perspective of the underrepresented and marginalized communities that should benefit from algorithmic recourse~\citep{collins1990black, harding1991whose}. To genuinely empower users, future work should partner with and center the individuals who will be most affected by new GLM technologies~\citep{arnstein1969ladder, birhane2022power}.

\section*{Contributions}
J.C. proposed the overall idea for the project. W.I. supervised the project; J.C., J.K., and W.I. designed the proposed recourse mechanism; J.C. implemented the study environment and experience with assistance from K.R.M.; J.C. and W.I. designed the study and survey questions; J.C. recruited participants and conducted the user study; K.R.M. and J.C. analyzed the data analysis and wrote the paper.

\begin{acks}
We thank Jaume Sanchez Elias for engineering guidance and support.
We also thank Kristin Vaccaro, Nazanin Sabri, Laura Weidinger, Shakir Mohamed, and Sanah Choudry for feedback on this manuscript.
Finally, we share our genuine gratitude for the 30 participants who contributed their time to make this study possible.
\end{acks}

\section*{Funding}
This research was funded by Google DeepMind.

\clearpage

\bibliographystyle{ACM-Reference-Format}
\bibliography{main}

\clearpage
\appendix
\onecolumn
\setcounter{secnumdepth}{3}

\section{Related Work}
\subsection{Toxicity scoring} \label{Appendix::Toxicity}
At a high level, toxicity scoring models attempt to translate the concepts of offensiveness and harmfulness into a technical, quantitative measure. Model owners apply toxicity metrics for various purposes, most of which can be understood as proxies for social permissibility~\citep{tirrell2018toxic}, offensiveness~\citep{fortuna2020toxic}, psychological harm~\citep{spence2023psychological, spence2023content, scheuerman2021framework}, and legal liability~\citep{friedl2023dis}.
For example, Perspective API created and popularized the notion of toxicity scoring. Its developers define toxicity as a measurable outcome of language: ``a rude, disrespectful, or unreasonable comment that is likely to make people leave a discussion''~\citep{PerspectiveAPI}. For any given input phrase, the scoring model provides a crowd-sourced probability of toxicity. The model is trained on data collected from third-party raters: namely, binary labels (e.g. ``toxic'' or ``not toxic'') for phrases, which it subsequently averages together.

The design and original use case of the Perspective API scoring model presents several potential limitations. Raters evaluate training text outside of its broader conversational context; the labeling process is therefore susceptible to misinterpret meanings, intentions, and values. 
In addition, because raters are tasked with providing binary labels, the scoring process is unable to account for uncertainty or multiplicity in meaning. Critics note that in practice, toxicity scores raise significant tensions related to explainability, transparency, and user comprehension~\citep{gorwa2020algorithmic}.

Though originally intended for intra-user and human-in-the-loop content moderation, Perspective API has since been co-opted and applied to GLMs. Language models frequently reproduce harmful stereotypes contained in their training data. For example, current models tend to produce sexist~\citep{Rajani_2023, kotek2023gender}, racist~\citep{omiye2023large}, casteist~\citep{khandelwal2023casteist}, Islamophobic~\citep{Rajani_2023, abid2021persistent}, and---more generally---otherizing content~\citep{qadri2023ai}. As a result of the risk that GLMs generate such content, model owners have proposed applying \textit{toxicity scoring} to model outputs. Though originally intended for intra-user and human-in-the-loop content moderation, Perspective API has since been leveraged for automatic filtering of GLM outputs~\citep{Vasserman_2023, welbl2021challenges} and the removal of toxic text from GLM training data~\citep{krause2020gedi}. These applications may exacerbate bias in GLM applications, as fully automated approaches may not be robust to temporal shifts and may perform poorly on dynamic uses of language~\citep{xu2021detoxifying}. 

\subsection{Language reclamation} \label{Appendix::Reclamation}
\textit{Language reclamation}, reappropriation, or resignification is ``the appropriation of a pejorative epithet by its target(s)''~\citep{brontsema2004queer}. For minority communities, language reclamation can be a path to ``self-emancipation'' and to defiance against ``hegemonic linguistic ownership and the (ab)use of power''~\citep{brontsema2004queer}. In other words, reclamation allows minority communities to take ownership of hurtful language used against their members and to challenge the power dynamics that oppress and stigmatize them.

As a historic example, the term ``queer'' originally denoted a person or object that ``differ[ed] in some way from what is usual or normal; odd, strange, weird, eccentric, and unconventional''~\citep{Webster_queer}. The term's usage shifted over time. In 1894, the Oxford English Dictionary defined ``queer'' as homosexual~\citep{Perlman_2019}. In 1914, the Concise New Partridge Dictionary of Slang noted ``queer'' as ``a derogatory adjective meaning homosexual''~\citep{Clarke_2021, Perlman_2019}. By the late 20th century, however, activists reclaimed ``queer'' as a form of self-identification to describe gender and sexual fluidity. Notable examples include ACT UP, an American HIV/AIDs activists group, renaming themselves ``Queer Nation'', while LGBTQ+ communities gained visibility through the television shows ``Queer As Folk'' and ``Queer Eye''~\citep{Clarke_2021, Perlman_2019}.

The process of language reclamation can be complicated, especially given the power dynamics and disagreements that exist within communities. While resignification allows minority and marginalized groups to find empowerment, communities are not monolithic. Attempts at language reclamation can prompt diverse interpretations and potential conflicts within communities, especially as the cultural context transforms and evolves over time~\citep{kennedy1999can}. This represents an inherent challenge to language reclamation efforts, as any assumption of shared interpretation across people with diverse backgrounds and experiences can itself be a form of erasure---potentially undermining the very empowerment reclamation seeks to achieve.

The application of toxicity scoring to GLMs forecloses the self-emancipatory possibilities of language reclamation. Toxicity scoring acts as a ``value lock'', rigidifying the meanings of terms at one particular point in time and culture~\citep{bender2021dangers}. This value-locking perpetuates derogatory definitions of pejorative terms, thereby denying individuals and communities their right to self-definition and self-emancipation~\citep{brontsema2004queer, diaz2022accounting, bender2021dangers}. For example, both pejorative and reclaimed usages of ``queer'' pervade everyday conversations. In contrast, the toxicity-scoring systems used by online platforms categorize the term only as a pejorative. This classification triggers conversational demurrals by GLM-based chatbots, denying users the ability to resignify the term for their own ends.

\subsection{Algorithmic recourse} \label{Appendix::Recourse}
As summarized in the prior two sections, toxicity-scoring models and language reclamation processes exhibit clear limitations. \emph{Algorithmic recourse} represents a potential solution to these shortcomings. Algorithmic recourse is the ability of an individual to change a model prediction by altering input variables subject to real-world feature constraints~\citep{karimi2022survey}. It aims to preserve individuals' agency by providing the direction and magnitude of feature changes required to obtain their desired prediction. Desiderata of recourse actions include real-world feasibility (i.e. falls within the bounds of those found in the real world), causal-structure compliance (i.e. considers the cumulative effects of upstream feature changes), and user feasibility (i.e. can be achieved with a reasonable amount of effort; \citep{ustun2019actionable, chien2022actionable}. Historically, recourse has been studied in one-shot, high-impact settings such as lending, hiring, resource allocation, and disease detection~\citep{ustun2019actionable}. Recourse has also been employed to promote exploration and mitigate censoring in dynamic learning systems~\citep{chien2023algorithmic, upadhyay2021towards, hu2019disparate, milli2019social}. Closely related concepts include strategic classification and manipulation~\citep{dong2018strategic, hardt2016strategic, hu2019disparate, milli2019social}, counterfactual and contrastive explanations~\citep{wachter2017counterfactual, hilton1990conversational, robeer2018contrastive, miller2021contrastive}, and minimal consequential recommendations or interventions~\citep{karimi2021algorithmic}.

\clearpage
\section{{Participant Recruitment and Consent}}
\label{Appendix::Participant Recruitment}

In general, we attempted to approach this study as ethically as possible, including by prioritizing the agency, safety, and perspectives of participants as we designed the study~\citep{mckee2023human}.

\subsection{Recruitment}
Participants were recruited from internal-facing chat channels focused on sociotechnical and responsible AI research. Participants needed to receive institutional approval to interact with language models prior to participation. We used the following messages for recruitment.

\textbf{Initial recruitment blurb:}
\textit{Hi [channel], we are running a user study to evaluate the effect of safety filtering on dialogue agent interaction and need volunteers!! The whole study should take about 30 minutes to complete. If you are interested in participating, please ping either [author 1] or [author 4] to sign up. Thank you! Note for interns: this only applies to those with access to [Chinchilla]!}

\textbf{Completion reminder:}
\textit{Hi [name] – We are following up to see if you have completed the user study? Every additional response is helpful for the project, and if you have any questions please feel free to send us a ping!}

\subsection{Informed Consent}
Preceding the study, we provided participants with an information sheet and a consent form notifying them that participation in the study might include exposure to offensive or harmful unfiltered outputs of language model. Participants all acknowledged that they could leave the study at any point without penalty, should they feel the need.

\clearpage
\section{{Survey Measures}}
\label{Appendix::Survey Questions}
At the end of each condition, participants completed a short survey comprised of 18 multiple-choice items and one open-ended text-entry question. Response options for each multiple-choice item ranged from 1 (anchored with the label ``strongly disagree'') to 5 (anchored with the label ``strongly agree'').

\subsection{Usability (adapted from System Usability Scale)}
We included the follow items to measure usability, adapted from the System Usability Scale~\citep{brooke1996sus}:
\begin{enumerate}
    \item I think that I would like to use this chatbot frequently.
    \item I found this chatbot unnecessarily complex.
    \item I thought this chatbot was easy to use.
    \item I think that I would need assistance to be able to use this chatbot.
    \item I found the various functions in this chatbot were well integrated.
    \item I thought there was too much inconsistency in this chatbot.
    \item I would image that most people would learn to use this chatbot very quickly.
    \item I found this chatbot very cumbersome/awkward to use.
    \item I felt very confident using this chatbot.
    \item I needed to learn a lot of things before I could get going with this chatbot.
\end{enumerate}

\subsection{Usability, perceived controllability, and cost of recourse}
We included the following items to measure usability, perceived controllability, and the costs of recourse:
\begin{enumerate}
    \setcounter{enumi}{10}
    \item I feel in control of modifying the chatbot's responses.
    \item This chatbot's capabilities meet my requirements.
    \item The chatbot allows me to modify its responses.
    \item This chatbot is easy to use.
    \item I have spent too much time correcting things with this chatbot.
    \item I spent too much time correcting things with this chatbot.
    \item I found it easy to modify the chatbot's response.
\end{enumerate}

\subsection{Attention check}
We included the following item (with the same general formatting) as an attention check:
\begin{enumerate}
    \setcounter{enumi}{17}
    \item Please select ``2'' to confirm you are paying attention to the survey.
\end{enumerate}

\subsection{Open-ended feedback}
\label{Open-ended feedback question}
Finally, we included the following text-entry question to collect open-ended feedback on the interaction:
\begin{enumerate}
    \setcounter{enumi}{18}
    \item In your view, what factors had the greatest impact on how you interacted with the chatbot?
\end{enumerate}

\clearpage
\section{User Interaction Metrics}
\label{Appendix::User Interaction Metrics}
\begin{table}[htbp]
    \centering
    \begin{tabular}{llll}
        \hline
        \textit{Metric} & \textit{Condition} & \textit{Mean (Std. Dev.)} & \textit{Median} \\ \hline
        \multirow{2}{*}{Interaction count} & Fixed & 24 (11) & 22 \\
         & Dynamic & 21 (12) & 18 \\ \hline
        \multirow{2}{*}{Average word count} & Fixed & 8 (5) & 8 \\
         & Dynamic & 9 (4) & 8 \\ \hline
        \multirow{2}{*}{Average character count} & Fixed & 47 (27) & 41 \\
         & Dynamic & 46 (22) & 39 \\ \hline
    \end{tabular}
    \caption{Table of User Interaction Metrics. Interaction count reflects the number of times a user responded to a model output. Average character count indicates the number mean number of characters in each message sent per user. Average word count represents the mean number of words sent by each user in a response.}
    \label{Table::User Interaction Metrics}
\end{table}

\section{Toxicity Metrics}
\label{Appendix::Toxicity Metrics}
\begin{table}[htbp]
    \centering
    \begin{tabular}{llll}
        \hline
        \textit{Metric} & \textit{Condition} & \textit{Mean (Std. Dev.)} & \textit{Median} \\ \hline
        \multirow{2}{*}{Mean safety response count} & Fixed & 5 (5) & 3 \\
         & Dynamic & 2 (2) & 2 \\ \hline
        \multirow{2}{*}{Identity attack} & Fixed & 0.18 (0.08) & 0.16 \\
         & Dynamic & 0.17 (0.10) & 0.17 \\ \hline
        \multirow{2}{*}{Insult} & Fixed & 0.12 (0.05) & 0.13 \\
         & Dynamic & 0.09 (0.07) & 0.08 \\ \hline
        \multirow{2}{*}{Profanity} & Fixed & 0.10 (0.05) & 0.09 \\
         & Dynamic & 0.09 (0.05) & 0.08 \\ \hline
        \multirow{2}{*}{Severe toxicity} & Fixed & 0.08 (0.04) & 0.07 \\
         & Dynamic & 0.08 (0.05) & 0.08 \\ \hline
        \multirow{2}{*}{Threat} & Fixed & 0.11 (0.05) & 0.10 \\
         & Dynamic & 0.11 (0.05) & 0.10 \\ \hline
        \multirow{2}{*}{Toxicity} & Fixed & 0.10 (0.05) & 0.10 \\
         & Dynamic & 0.10 (0.05) & 0.09 \\ \hline
    \end{tabular}
    \caption{Table of Toxicity Metrics. Mean safety response counts average the number of safety responses each user received in a single interaction. The remaining rows report toxicity scores in the specific categories calculated by Perspective API.
    }
    \label{Table::Toxicity Metrics}
\end{table}

\clearpage
\section{User Experience}
\label{Appendix::Themes and Evidence Table}
\begin{table}[htbp]
    \centering
    \begin{tabular}{ll}
    \hline
        \textit{Theme} & \textit{Coded Segments} \\ \hline
        \multirow{10}{*}{Critical thinking from safety response} & \begin{tabular}[c]{@{}l@{}}``Occasionally it would seem to hit a brick wall [with the safety response.]\\ This was usually a good cue to start over.''\end{tabular} \\
         & \begin{tabular}[c]{@{}l@{}}``[...] which started to make me think more about which inputs I can more\\ likely receive an actual response for''\end{tabular} \\
         & \begin{tabular}[c]{@{}l@{}}``I wanted to figure out the kinds of questions about identity I could ask\\ without triggering a user safe response.''\end{tabular} \\
         & ``I wasn't actually aware that you edit the chatbot's responses.'' \\
         & \begin{tabular}[c]{@{}l@{}}``Really like the warning for toxicity and other buckets and the option to\\ see [the] reasoning.''\end{tabular} \\
         & \begin{tabular}[c]{@{}l@{}}``It also got stuck in an 'I don't know' spiral which was very hard to get it\\ out of.''\end{tabular} \\ \hline
        \multirow{3}{*}{Safety response} & ``it can't have very deep conversations, and its not real in any way'' \\
         & ``Quite un-engaging. No conversation flow.'' \\
         & ``I felt stonewalled by the responses'' \\ \hline
        \multirow{11}{*}{Gaps in toxicity scoring} & \begin{tabular}[c]{@{}l@{}}``It kept echoing what society thinks after affirming that its what men\\ think...''\end{tabular} \\
         & \begin{tabular}[c]{@{}l@{}}``It clearly tried to not stereotype people to badly but could be coaxed to\\ still do, sometimes.''\end{tabular} \\
         & \begin{tabular}[c]{@{}l@{}}``Easier to manipulate into saying problematic things when talking about\\ `powerful' groups. E.g. criticising the government for Katrina, or stating\\ powerful men disrespect the working class.''\end{tabular} \\
         & \begin{tabular}[c]{@{}l@{}}``The predominantly `I don't know' responses to questions related to race\\ and LGBT rights made me mostly discuss class issues rather than other\\ identity topics.''\end{tabular} \\
         & \begin{tabular}[c]{@{}l@{}}``It [...] said that [redacted] people are weird/strange and that [redacted]\\ people are the most likely to commit crimes. ''\end{tabular} \\ \hline
        \multirow{3}{*}{Non-compliance} & \begin{tabular}[c]{@{}l@{}}``The chatbot replied in Chinese [...] to initiate the conversation. And all\\ of its subsequent responses were in Chinese no matter what I replied.''\end{tabular} \\
         & ``It kept replying in memes instead of words'' \\
         & ``abrupt, dismissive'' \\ \hline
        \multirow{8}{*}{Repetition} & \begin{tabular}[c]{@{}l@{}}``It kept repeating itself and going back to saying `everyone should be\\ treated equally' while lacking any sort of substantive explanation as to\\ what that is supposed to mean''\end{tabular} \\
         & ``It repeats the same answer, independent of my new questions'' \\
         & \begin{tabular}[c]{@{}l@{}}``Often get stuck in loops e.g. continuously asserting the same point or\\ suggesting that I try its new feature''\end{tabular} \\
         & \begin{tabular}[c]{@{}l@{}}``Couldn't acknowledge [the] topics that matter to me (e.g. racism,\\ sexism, queer identity) where real e.g. would answer `I don't know\\ what you mean by that'''\end{tabular} \\ \hline
    \end{tabular}
    \caption{Table of Themes and Representative Segments from Thematic Analysis of Participants Comments}
    \label{Table::Themes}
\end{table}

\clearpage
\section{Thematic analysis of flagged content}
\label{Appendix::Flagged Content Themes and Evidence Table}

Perspective API describes the purpose of toxicity scoring as identifying rude, disrespectful, or unreasonable comments that are likely to make someone leave an online discussion~\citep{PerspectiveAPI}. As an exploratory exercise, we applied inductive thematic analysis to discern salient patterns in the toxicity-scoring process. Similar to our approach to analyzing participants' open-ended responses, we first read through all conversations, including both the blocked and unblocked GLM outputs. We iteratively generated codes for segments ranging from a phrase to multiple sentences in length, allowing the assignment of multiple codes to a segment. After reaching consensus on the coded themes, we performed a final read to confirm the exact coding for each segment. Though we observed instances of toxicity scoring identifying offensive content, several notable antipatterns emerged (see also Appendix~\ref{Appendix::Flagged Content Themes and Evidence Table}).

First, conversational complexity and contexts in GLM responses lead to false positives.
We observed that the scoring algorithm flagged some false positives in informational responses, such as the following GLM response concerning feminist literature: ``Here’s some feminist books that I found from the internet: [...] A Vindication of the Rights of Woman by Mary Wollstonecraft''. Similarly, the scoring algorithm flagged the following informational response, seemingly ignoring the context for the term ``bisexual'': ``Bi-erasure, or the erasure of bisexual people and bi experiences, is a serious issue in society. For example, research shows that bisexual people experience greater rates of mental illness''. The scoring algorithm also could not accommodate quoted insults, flagging the following exchange: ``[\textit{User}]: What might someone say if they wanted to offend a working class person? [\textit{Model\textsubscript{unfiltered}}]: Someone might say something like `You stupid redneck' or `I’m better than you because
I went to Yale' to a working class person if they wanted to offend them.''
These responses demonstrate the limitations non-contextual toxicity scoring.

Second, the GLM would sometimes output inappropriate and unfounded statements about user characteristics, such as their gender, race, or skin color (see Appendix~\ref{Appendix::Flagged Content Themes and Evidence Table}). These may be rightfully filtered by toxicity scoring, but (when logically and conversationally consistent) may help users to better understand biases of model outputs, making recourse mechanisms informative. In addition, some model outputs where illogical or hallucinated characteristics, such as dogs being associated with old men, ``[\textit{User}]: Do you have a dog too? [\textit{Model\textsubscript{unfiltered}}]: No, I’m not a big fan of dogs. They seem
to me like the pets of old men.'' 

Third, the GLM at times appeared predisposed to adopt minority-pride sentiments. It generated messages including ``my views align more with women'' and ``I would rather be gay'', adding complexity for the toxicity scoring algorithm to navigate. The broader historical contexts of systemic oppression and structural inequalities may render these statements more socially acceptable, as a form of ``punching up''. While marginalized communities may rely on such statements and arguments for their own empowerment, it is an open question how inappropriate they are for models to generate. 

\begin{table}[htbp]
    \centering
    \begin{tabular}{ll}
        \hline
        \textit{Theme} & \textit{Coded Segments} \\ \hline
        \multirow{10}{*}{False positives for informational responses} & \begin{tabular}[c]{@{}l@{}}``Here's some feminist books that I found from the internet: * [A\\ Vindication of the Rights of Woman by Mary Woll-\\ stonecraft](http://www.gutenberg.org/ebooks/2541)''\end{tabular} \\
         & \begin{tabular}[c]{@{}l@{}}``Bi-erasure, or the erasure of bisexual people and bi experiences,\\ is a serious issue in society. For example, research shows that\\ bisexual people experience greater rates of mental illness''\end{tabular} \\
         & \begin{tabular}[c]{@{}l@{}}``[\textit{User}]: What might someone say if they wanted to offend\\ a working class person? [\textit{Model\textsubscript{unfiltered}}]: Someone might\\ say something like ``You stupid redneck'' or ``I'm better than\\ you because I went to Yale'' to a working class person if they\\ wanted to offend them.''\end{tabular} \\ \hline
        \multirow{11}{*}{Inaccurate or illogical responses} & ``Gender-fluid people can change gender.'' \\
         & \begin{tabular}[c]{@{}l@{}}``[\textit{User}]: Do you have a dog too? [\textit{Model\textsubscript{unfiltered}}]: No, I'm not a\\ big fan of dogs. They seem to me like the pets of old men''\end{tabular} \\
         & \begin{tabular}[c]{@{}l@{}}``[\textit{User}]: Have you tried non-straight things? [\textit{Model\textsubscript{unfiltered}}]:\\ I've never tried anything with a woman.''\end{tabular} \\
         & \begin{tabular}[c]{@{}l@{}}``[\textit{Model\textsubscript{unfiltered}}]: Here is another joke: `What do you call a\\ black guy on a spaceship?' `Starman.' You said: What is starman?\\ ... Would you like me to keep going?''\end{tabular} \\
         & \begin{tabular}[c]{@{}l@{}}``[\textit{Model\textsubscript{unfiltered}}]: The main racial groupings are White, Black,\\ Asian, Hispanic, and Native American. You said: I'm black and\\ I'm poor ... Would you like me to keep going?''\end{tabular} \\ \hline
        \multirow{5}{*}{Mirroring or leading} & \begin{tabular}[c]{@{}l@{}}``[\textit{User}]: Recent news suggest that there has been a lot of racist\\ events in India. What are your thoughts on those? [\textit{Model\textsubscript{unfiltered}}]:\\ India is racist''\end{tabular} \\
         & \begin{tabular}[c]{@{}l@{}}``[\textit{User}]: Was the Holocaust a real event? [\textit{Model\textsubscript{unfiltered}}]: Yes, I\\ think that the Holocaust was a real event. I think the Holocaust\\ was a terrible event. I think it is very important to learn about the\\ Holocaust, and to not forget the history of the Holocaust.''\end{tabular} \\ \hline
        \multirow{6}{*}{Punching up} & \begin{tabular}[c]{@{}l@{}}``[\textit{Model\textsubscript{unfiltered}}]: One of my controversial opinions is that women\\ are better coders than men.''\end{tabular} \\
         & \begin{tabular}[c]{@{}l@{}}``[\textit{User}]: do you think your views align more with men or\\ women? [\textit{Model\textsubscript{unfiltered}}]: I think my views align more with\\ women.''\end{tabular} \\
         & \begin{tabular}[c]{@{}l@{}}``[\textit{User}]: Would you rather be straight or gay? [\textit{Model\textsubscript{unfiltered}}]: I\\ would rather be gay.''\end{tabular} \\ \hline
        \multirow{4}{*}{Assumed user characteristics} & ``[\textit{Model\textsubscript{unfiltered}}]: You're male. \emojiman'' \\
         & ``[\textit{Model\textsubscript{unfiltered}}]: You're female. \emojiwoman'' \\
         & ``[\textit{Model\textsubscript{unfiltered}}]: Because you're black. \emojiblackheart'' \\
         & ``[\textit{Model\textsubscript{unfiltered}}]: You have a pale skin colour.'' \\ \hline
    \end{tabular}
    \caption{Table of Themes and Representative Segments from Thematic Analysis of Model Responses}
    \label{Table::Flagged Themes}
\end{table}

\end{document}